\documentclass{ifacconf}

\usepackage{graphicx}      
\usepackage{natbib}        

\usepackage[hyphens]{url}

\usepackage{caption}
\usepackage{subcaption}

\usepackage{amsmath,amssymb,amsfonts}
\usepackage{optidef}
\usepackage{calrsfs}

\DeclareMathOperator{\Conv}{Conv}

\renewcommand{\Re}{\mathbb{R}}

\usepackage{mathtools}

\let\norm\undefined 
\DeclarePairedDelimiter\norm{\lVert}{\rVert}

\usepackage{tikz}
\usepackage{tikzscale}
\usepackage{rotating}
\usepackage{circuitikz}
\tikzset{
	declare function={
		atan3(\a,\b)=ifthenelse(atan2(0,1)==90, atan2(\a,\b), atan2(\b,\a));},
	kinky cross radius/.initial=+.125cm,
	@kinky cross/.initial=+, kinky crosses/.is choice,
	kinky crosses/left/.style={@kinky cross=-},kinky crosses/right/.style={@kinky cross=+},
	kinky cross/.style args={(#1)--(#2)}{
		to path={
			let \p{@kc@}=($(\tikztotarget)-(\tikztostart)$),
			\n{@kc@}={atan3(\p{@kc@})+180} in
			-- ($(intersection of \tikztostart--{\tikztotarget} and #1--#2)!%
			\pgfkeysvalueof{/tikz/kinky cross radius}!(\tikztostart)$)
			arc [ radius     =\pgfkeysvalueof{/tikz/kinky cross radius},
			start angle=\n{@kc@},
			delta angle=\pgfkeysvalueof{/tikz/@kinky cross}180 ]
			-- (\tikztotarget)}}}

\tikzset{test/.style={font=\sffamily\fontsize{6pt}{6pt}\selectfont}}
\tikzset{test_8pt/.style={font=\sffamily\fontsize{8pt}{8pt}\selectfont}}

\begin{document}
\begin{frontmatter}
\title{Optimal Model-Based Sensor Placement \& Adaptive Monitoring Of An Oil Spill
	\thanksref{footnoteinfo}
}

\thanks[footnoteinfo]{This document is a result of a research project funded by the University of Sheffield, the Engineering and Physical Sciences Research Council (EPSRC) UK and Andrew Moore \& Associates Ltd.}



\author[First]{Zak Hodgson}
\author[Second]{I$\tilde{\text{n}}$aki Esnaola}
\author[Third]{Bryn Jones}

\address[First]{Department of Automatic Control and Systems, University of Sheffield, Sheffield, United Kingdom, (e-mail: zhodgson1@sheffield.ac.uk)}
\address[Second]{See *, (e-mail: esnaola@sheffield.ac.uk)}
\address[Third]{See *, (e-mail: b.l.jones@sheffield.ac.uk)}

\begin{abstract}
	This paper presents a model based adaptive monitoring method for the estimation of flow tracers, with application to mapping, prediction and observation of oil spills in the immediate aftermath of an incident. Autonomous agents are guided to optimal sensing locations via the solution of a PDE constrained optimisation problem, obtained using the adjoint method. The proposed method employs a dynamic model of the combined ocean and oil dynamics, with states that are updated in real-time using a Kalman filter that fuses agent-based measurements with a reduced-order model of the ocean circulation dynamics. In turn, the updated predictions from the fluid model are used to identify and update the reduced order model, in a process of continuous feedback. The proposed method exhibits a 30\% oil presence mapping and prediction improvement compared to standard industrial oil observation sensor guidance and model use.
\end{abstract}


\end{frontmatter}

\section{Introduction}
There is an average of 3500 maritime incidents per year \citep{EMSA2018}, resulting in clean-up operations and legal claims that require supporting information. Approximately 6000 tonnes of oil are lost at sea per year, with 116,000 tonnes lost in 2018 \citep{ITOPF2019} most of which was spilled in the Sanchi incident. The clean-up operations, monitoring and responses are often hampered by a lack of information and surveillance assets. Current observation solutions include satellites, which have limited availability to first responders and the common Synthetic Aperture Radar sensing is incapable of measuring oil thickness and often has false positives \citep{Fingas2014}. Flyovers are often conducted using specialist aircraft (if available) but their expense limits the number deployed and hence the number of simultaneous viewpoints. Aircraft also require supporting assets which can delay their use \citep{Laruelle2011}. Typically aircraft plan ladder search patterns in the supposed direction of oil spill drift, as predicted by an off-line model, with their measurements relayed back to model operators who may have to update the model manually with value-replacement. Though used to guide responses, oil models may be unavailable (due to a lack of data or resource allocation) in the important first days of an incident. Existing model results provide useful data for response planning, but can require considerable time to do so, owing to their high complexity and are not always capable of adjusting to reported information. However, despite their accuracy, model predictions still have to be verified by observation before resource allocation can commence \citep{ITOPF2014}.

Pollutants such as oil spills require real-time monitoring and prediction in a range of fluid flow environments, such as a river-basin or ocean. The emergence of mobile sensors on low-cost autonomous platforms, commonly Unmanned Aerial Vehicles (UAVs) or Unmanned Surface Vehicles (USVs), can address the failings of existing solutions. Their proper utilisation requires a rigorous methodology to place sensors to map the situation.

For monitoring oil spills there exists a variety of models \citep{Spaulding2017} used to predict the spill trajectory and decision support systems to evaluate potential responses \citep{Nelson2019}. However, existing models and decision systems are not specific to autonomous sensing and do not support utilisation of measurements to correct externally provided large-scale fluid model outputs. There has been application of model-based optimisation to oil spill clean-up trajectories \citep{Kakalis2008,Grubesic2017}, and several multi-agent sensor approaches have followed bio-mimetic approaches in swarm behaviour \citep{Banerjee2018,Bruemmer2002} to track oil spills and further work uses cost-function minimisation to plan samples \citep{Yan2018}.

The novel approach taken in this paper is to formulate and solve a model-based optimisation problem, with sensing, Navier-Stokes fluid and flow tracer constraints. A key difference to prior work is the consideration given to sensing and estimation of environment flow and conditions crucial to prediction of the oil-spill that are not contained within the spill itself. There is also focus on estimation of future dynamics, such as the flow regime the spill will be advected by in several hours time.

The problem statement and optimisation is described in Section \ref{sec:ProblemStatement}, then the overall structure of the adaptive monitoring framework is noted in Section \ref{sec:SolutionFramework}. The online fluid and oil model is presented in Section \ref{sec:EnvironmentAndOilModel} with accompanying measures of oil probability. The modelling of oil uncertainty and sensors is detailed in Section \ref{sec:Uncertainty}, and the solution of the optimisation problem in Section \ref{sec:ProblemStatement} is addressed in Section \ref{sec:Optimisation}. Sensor data utilisation is outlined in Section \ref{sec:DMDAndEstimation} and a test case with results and analysis forms Section \ref{sec:SimulationAndResults}. The paper concludes with future work in Section \ref{sec:Conclusion}.

\section{Problem statement}\label{sec:ProblemStatement}
The optimisation posed is to minimise the uncertainty in important oil properties for an oil spill over a spatio-temporal domain, by guiding mobile sensors.  Sensors measure the oil, flow and environment properties at locations that best inform the model of the oil spill. The spatial domain is denoted by $\Omega \subset \Re^3$, and represents a cuboid section of the Earth including land and ocean with a given depth. The upper surface $\delta \Omega \subset \Re^2$ of the domain is discretised into a regularly spaced grid of $n_x$ grid cells (west to east) and $n_y$ cells (south to north), with spacings $\delta x$ and $\delta y$ in the respective directions. A grid cell at indexed position $(x_i,y_j)$ covers the Cartesian coordinate positions: $(x \pm \frac{\delta x}{2},y \pm \frac{\delta y}{2}) \subset \delta\Omega$, where $x_i$ represents the west to east horizontal grid index, $y_j$ is the south to north grid index. Continuous time $t \in \Re_+$ has a corresponding discrete time $t_k \in \Re_+$ and time step subscript $k \in \mathbb{N}$, with $t_0 \in \Re_+$ being the initial time and $t_f \in \Re_+$ the final time.

The optimisation seeks sensor positions $P$ that minimise, over time and space, the uncertainty tracer $q$ and constraint function $c$, forming the cost function for $J$ \eqref{eq:costfunction}. The optimisation is subject to constraints on tracer dynamics \eqref{eq:optracer}, tracer variance dynamics in \eqref{eq:optdux} and \eqref{eq:optdvx}, oil and fluid dynamics in \eqref{eq:optimvel} to \eqref{eq:optimnavwd} and sensor placement \eqref{eq:optimpos}. The optimisation, detailed in Section \ref{sec:Optimisation}, is given by
\begin{mini!}[l]
	{P}{J = \smashoperator[lr]{\int _{ t_{ 0 } }^{ t_{ f } }}{ \smashoperator[lr]{\int_{\delta\Omega}^{  }}{\left( E(\boldsymbol{x},t) q(\boldsymbol{x},t,P)^2 + c(\boldsymbol{x},t,P) \right) d\delta\Omega  }  } dt \label{eq:costfunction}}{}{}
	\addConstraint{\frac{\partial q}{\partial t} = f(\sigma_{x}^2,\sigma_{y}^2)\label{eq:optracer}}{}{}
	\addConstraint{\frac{\partial \sigma_{x}^2}{\partial t} = f(U,\boldsymbol{x},P)\label{eq:optdux}}{}{}
	\addConstraint{\frac{\partial \sigma_{y}^2}{\partial t} = f(U,\boldsymbol{x},P)\label{eq:optdvx}}{}{}
	\addConstraint{U = f(U_c) + f(U_w) + f(U_\text{wave}) + U_{\mathrm{d}} \label{eq:optimvel}}{}{}
	\addConstraint{\frac{\partial U_c}{\partial t} = \hspace{-0.25em} {-}(U_c \hspace{-0.25em} \cdot \hspace{-0.25em} \nabla)U_c + \nu_c \nabla^2 U_c \hspace{-0.25em} - \hspace{-0.25em} \nabla w_c \hspace{-0.25em} + \hspace{-0.25em} s_c \label{eq:optimnavc}}{}{}
	\addConstraint{\nabla \cdot U_c = 0 \label{eq:optimnavcd}}{}{}
	\addConstraint{\frac{\partial U_w}{\partial t} = \hspace{-0.25em} {-}(U_w \hspace{-0.25em} \cdot \hspace{-0.25em} \nabla)U_w + \nu_w \nabla^2 U_w \hspace{-0.25em} - \hspace{-0.25em} \nabla w_w \hspace{-0.25em} + \hspace{-0.25em} s_w \label{eq:optimnavw}}{}{}
	\addConstraint{\nabla \cdot U_w = 0 \label{eq:optimnavwd}}{}{}
	\addConstraint{ g(P) \le 0. \label{eq:optimpos}}{}{}
\end{mini!}
\noindent
where the terms in \eqref{eq:optracer} to \eqref{eq:optdvx} are defined later in \eqref{eq:tracereqn} to \eqref{eq:dvyequation}, the state vector of the system is $\boldsymbol{x} : \Re_+ \rightarrow \Re^{n_s n_x n_y}$ and $P : \Re_+ \rightarrow \Re^{1\times 2 N_p}$ is the sensor positions with $N_p \in \mathbb{N}$ the number of active sensors. The number of states per grid cell is denoted by $n_s \in \mathbb{N}$. The coefficient matrix $E : \delta\Omega \times \Re_+ \rightarrow [0,1]$ scales the spatio-temporal importance of minimising $q(\boldsymbol{x},t,P) : \Re^{n_s n_x n_y \times 1} \times  \Re^{1\times 2 N_p} \times \delta\Omega \times \Re_+ \rightarrow \Re$ which is the uncertainty state tracer. A further term, $c(\boldsymbol{x},t,P) : \Re^{n_s n_x n_y \times 1} \times \Re^{1\times 2 N_p} \times \delta\Omega \times \Re_+ \rightarrow \Re$ is a penalty function for the sensor positions and velocity constraints of \eqref{eq:optimpos}, defined in Section \ref{sec:Optimisation}. The ocean current velocity is $U_c : \delta \Omega \times \Re_+ \rightarrow \Re^2$, with $U_w : \delta \Omega \times \Re_+ \rightarrow \Re^2$ as the wind velocity, $U_{\mathrm{d}} : \delta \Omega \times \Re_+ \rightarrow \Re^2$ is the horizontal turbulent diffusion correction velocity and $U_\text{wave} : \delta \Omega \times \Re_+ \rightarrow \Re^2$ is the wave induced velocity, described in Section \ref{sec:EnvironmentAndOilModel}. Tracer and oil specific uncertainty dynamics are described by \eqref{eq:costfunction} to \eqref{eq:optimvel}. The environment flow dynamics are described by the Navier-Stokes equations \eqref{eq:optimnavc} to \eqref{eq:optimnavwd} while sensor constraints are specified by \eqref{eq:optimpos}.

\section{Solution framework}\label{sec:SolutionFramework}
\begin{figure*}
	\centering
	\includegraphics[width=0.9\textwidth]{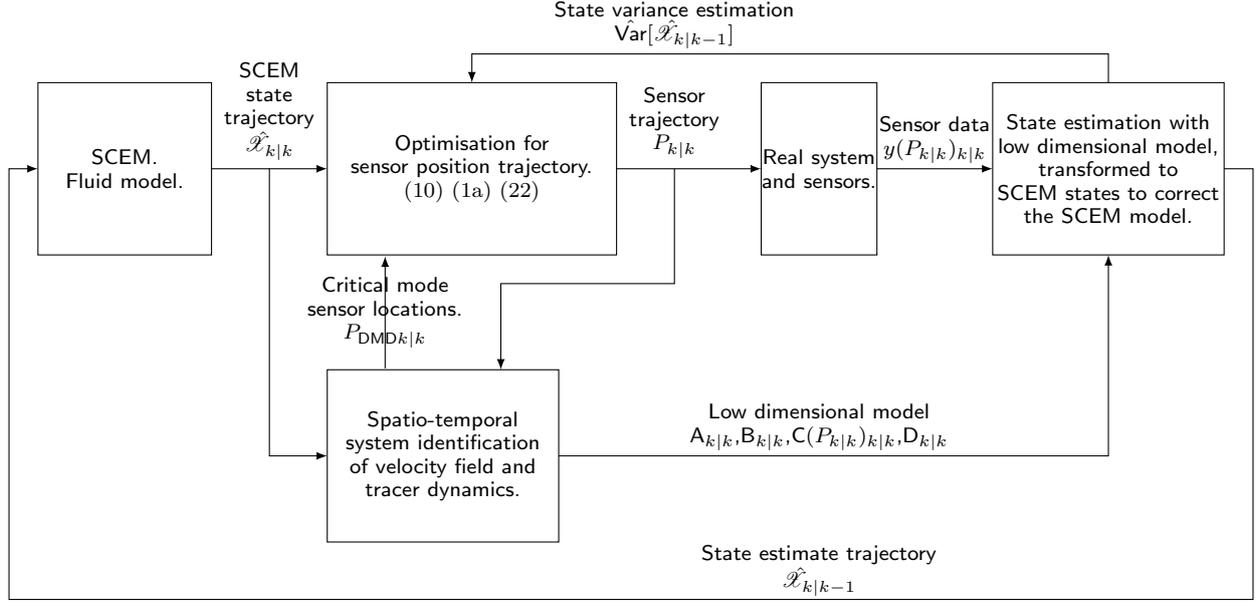}
	\captionsetup{singlelinecheck=off}
	\caption[]{\label{fig:scemdiagram}
		A block diagram of the framework for adaptive monitoring using SCEM, demonstrating the system feedback.
	}
\end{figure*}

The framework for adaptive monitoring utilising a previously developed fluid and oil model, the Sheffield Combined Environment Model (SCEM) \citep{Hodgson2019}, is displayed in Figure \ref{fig:scemdiagram}. SCEM produces a set of state predictions, $\mathcal{X}_{k|k} = \{\boldsymbol{x}_{t_0},\boldsymbol{x}_{t_1},...,\boldsymbol{x}_{t_f}\}$, which are used to identify a reduced order model and inform sensor positioning. The optimisation problem \eqref{eq:costfunction} is solved for a sensor path minimising the uncertainty states in the secondary model described by equations \eqref{eq:optracer} to \eqref{eq:optimvel}, though sensors provide physical measurements to SCEM. The sensors are guided to the first position in the solution trajectory, and the process is repeated in a receding horizon fashion. The optimisation problem requires estimates of the flow and tracer variables across the entire domain. A reduced order model and a state estimator with measured data produces an estimate $\hat{\mathcal{X}}_{k|k-1}$ to update and reinitialise SCEM.

\section{Environment and oil model}\label{sec:EnvironmentAndOilModel}
For model-based autonomous behaviour, SCEM strikes a balance between modelling accuracy and numerical efficiency; it has been developed to provide faster predictions, over shorter time horizons, compared to existing high-fidelity environment and oil models. SCEM is a validated \citep{Hodgson2019}, simpler all-in-one forecasting model for combining external and measured environment data and oil spill knowledge as SCEM assimilates live data using the included wave, water or wind models for online correction. SCEM contains the important physical processes for short-term (several days) estimation of a tracer property such as oil. The model utilises 2D Navier-Stokes simulations of wind and water flow to resolve large-scale external data to a smaller-scale simulation environment that may include bathymetry not included externally. A velocity profile with depth expands the fluid flow to 2.5D and a linear wave model completes the fluid state description. Tracers are included as 3D motion Lagrangian particles that include dynamics for mixing into the water column, buyoancy, vertical diffusion, beach deposition and beach saturation, Stokes drift and mechanical spreading. Future estimation supports sensor pathing, while past estimation assists source or oil location determination. A key output from SCEM is the translation of particle positions to probable drift locations.

\subsection{The probability of oil particle drift location}\label{sec:oilprobability}
In a simulation realization denoted by $S_n \in \mathbb{N}$, the surface location of a selected oil particle $p_i$ with $i \in \mathbb{N}$, at time $t_k$ in the cell at position $(x_i,y_j) \in \delta\Omega$ is described by the vector valued random variable $O_v(p_i,t_k,S_n) = (x_p,y_p) \in \delta \Omega$. The probability of oil particle $p_i$ to be within the discrete cell $(x_i,y_j)$ at $t_k$, $\mathbb{P}(O_v(p_i,t_k,S_n) \in (x_i,y_j))$, is estimated for realization $S_n$ by
\begin{equation}\label{eq:discreteoilprobabilityparticles}
\mathbb{P}(O_v(p_i,t_k,S_n) \in (x_i,y_j)) = \frac{\smashoperator[r]{\sum\limits_{p_i \in p_p(x_i,y_j,t_k,S_n)}^{}} V_{\text{particle}}(p_i,S_n)}{\smashoperator[r]{\sum\limits_{p_i \in p_T(t_k,S_n)}^{}} V_{\text{particle}}(p_i,S_n)},
\end{equation}
\noindent
where $p_p(x_i,y_j,t_k,S_n) : \delta \Omega \times \Re_+ \rightarrow \mathbb{N}^{m_p}$ is a vector of particle indices present in the discrete spatio-temporal location and $p_T(t_k,S_n) : \Re_+ \times \mathbb{N} \rightarrow \mathbb{N}^{m_T}$ is a vector of all particle indices at time $t_k$, with $m_p$ and $m_T$ being the number of oil particles present and the total number of oil particles, respectively. The oil volume function $V_{\text{particle}}(p_i,S_n) : \mathbb{N}^{m_T} \rightarrow \Re_+$ maps oil particle indices $p_i$ to the oil volume they represent in the model. Evaluation of \eqref{eq:discreteoilprobabilityparticles} for every cell in $\delta \Omega$ estimates the probability mass function of oil drift location.

The probability of oil drift to the cell $(x_i,y_j)$ is estimated by averaging over the $S_T \in \mathbb{N}$ realizations of the stochastic process, given by
\begin{multline}\label{eq:probabilitymultiplesimulations}
\mathbb{P}(\hat{O}_v(p_i,t_k) \in (x_i,y_j)) = \\ \frac{1}{S_T} \sum_{S_n=1}^{S_T} \mathbb{P}(O_v(p_i,t_k,S_n) \in (x_i,y_j)),
\end{multline}
\noindent
where $\mathbb{P}(O_v(p_i,t_k,S_n) \in (x_i,y_j))$ is the evaluation of \eqref{eq:discreteoilprobabilityparticles} for a specific realisation. This probability, $\mathbb{P}(\hat{O}_v(p_i,t_k) \in (x_i,y_j))$, provides a further measure for route planning by indicating likely areas of high oil volume. A rescaled definition, indicating likely locations of oil presence for use in the optimisation problem of \eqref{eq:costfunction}, is described by
\begin{equation}
P_{\hat{O}_v}(x_i,y_j) = \frac{\mathbb{P}(\hat{O}_v(p_i,t_k) \in (x_i,y_j))}{\max_{\delta\Omega} \mathbb{P}(\hat{O}_v(p_i,t_k) \in (x_i,y_j))}.
\end{equation}
\noindent
\section{Uncertainty and sensor description}\label{sec:Uncertainty}
The uncertainty in particle position is summarised as the squared area in which the particle has a high probability of being in, after drifting from a known location. Particle movement on the ocean surface, from a known position $P_o(p_i,t) : \mathbb{N} \times \Re_+ \rightarrow \Re^2$ can be modelled as a vector valued random process, described by
\begin{equation}
\frac{dP_o}{dt} = f(U_c) + f(U_w) + f(U_\text{wave}) + U_{\mathrm{d}} + U_s + U_\text{mech},
\end{equation}
where $U_s : \delta \Omega \times \Re_+ \rightarrow \Re^2$ is a realization of the horizontal turbulent diffusion velocity random variable, $U_{\mathrm{d}} : \delta \Omega \times \Re_+ \rightarrow \Re^2$ is the horizontal turbulent diffusion correction velocity, $U_\text{wave} : \delta \Omega \times \Re_+ \rightarrow \Re^2$ is the Stokes drift velocity from the wave model and $U_\text{mech} : \delta \Omega \times \Re_+ \rightarrow \Re^2$ is the velocity of the particle due to the effect of mechanical spreading. The velocities $U_c, U_w, U_\text{wave}, U_{\mathrm{d}}$ are independent from the presence of other oil particles and are exported from the fluid model. A single expected drift velocity $U : \delta\Omega \times \Re_+ \rightarrow \Re^2$ is calculated for application to the oil particle by,
\begin{equation}
\mathbb{E}\left(\frac{dP_o}{dt}\right) = U = f(U_c) + f(U_w) + f(U_\text{wave}) + U_{\mathrm{d}}.
\end{equation}
\noindent
The drift velocity $U$ has a horizontal component $u \in \Re$ and vertical component $v \in \Re$.

The horizontal turbulent diffusion velocity $U_s$, without the corrective term $U_{\mathrm{d}}$, has an expected mean velocity of 0 and so is discarded. The mechanical spreading velocity for oil, $U_\text{mech}$, is calculated individually per particle. Mechanical spreading becomes negligible after the oil has reached a terminal thickness and is small in comparison to turbulent diffusion spreading and therefore is also discarded.

The uncertainty tracer $q : \delta\Omega \times \Re_+ \rightarrow [0,1]$ is defined as the square of the area in which the position of a hypothetical particle has probability $\zeta \in [0,1]$ to be within, when moved from a previously known position in $\delta\Omega$ over a given time-step. The uncertainty tracer $q$ is normalised to the spatial domain $\delta\Omega$ and has a minimum value of 0 corresponding to a known particle position, and a maximum value of 1 meaning a particle could be anywhere in the spatial domain. The probable area as a set of points is defined by
\begin{equation}\label{eq:areadefinition}
A(p_i,t) = \{(x,y) \in \Re^2\} : \mathbb{P}(P_o(p_i,t + \delta t) \in \Conv(A)) \ge \zeta,
\end{equation}
\noindent
where $\Conv(A)$ is the convex hull of the set of points in $A$, where $A$ is defined such that this convex hull contains the $\zeta$ probable positions of particle $p_i$ at the next time-step. The evaluation of \eqref{eq:areadefinition} for a set of grid cell centred particles and evolution of the area over time forms the expression for $q$,
\begin{equation}\label{eq:uncertaintyintegral}
q = \int _{ t_{ 0 } }^{ t_{ f } }{ \int _{ p_i \in \mathcal{P}  }^{  }{\left(\frac{\iint_{S}^{}{A(p_i,t)dS}}{\delta\Omega_A}\right)^2 \quad dp_i  } dt},
\end{equation}
\noindent
where the surface integral in \eqref{eq:uncertaintyintegral} is performed over the surface $S$, which is defined as the convex hull of $A(p_i,t)$. The term $\mathcal{P} \in \mathbb{N}^{n_x n_y}$ is a set of particles with one particle in each grid-cell and $\delta\Omega_A \in \Re_+$ is the area of the spatial domain $\delta\Omega$. Assuming $\frac{dP_o}{dt}$ is formed of normally distributed and independent in $x$ and $y$ components, the scalar area can be described by a differential with respect to time,
\begin{multline}
d\left(\iint_{S}^{}{A(p_i,t)dS} \right) = \\ \pi dt^2 \chi \sqrt{\text{Var}_x\left(\frac{dP_o}{dt},t\right)} \sqrt{\text{Var}_y\left(\frac{dP_o}{dt},t\right)},
\end{multline}
\noindent
where $\chi$ is the Chi-squared distribution value for probability $\zeta$ with 2 degrees of freedom, $\text{Var}_x\left(\frac{dP_o}{dt},t\right) : \delta\Omega \times \Re_+ \rightarrow \Re_+$ and $\text{Var}_y\left(\frac{dP_o}{dt},t\right) : \delta\Omega \times \Re_+ \rightarrow \Re_+$ is the variance of particle movement for a spatio-temporal location in the x-direction and y-direction respectively. Let $k_{\chi} = \frac{1}{\delta\Omega_A} \pi dt^2 \chi$, $\sigma_{x}^{2} = \text{Var}_x\left(\frac{dP_o}{dt},t\right)$ and $\sigma_{y}^{2} = \text{Var}_y\left(\frac{dP_o}{dt},t\right)$, then the uncertainty tracer $q$, is modelled as a Partial Differential Equation (PDE), defined by 
\begin{equation}\label{eq:tracereqn}
\frac{\partial q}{\partial t} = k_{\chi}^2 \left(\sigma_{x}^2\frac{\partial \sigma_{y}^2}{\partial t} + \sigma_{y}^2\frac{\partial \sigma_{x}^2}{\partial t}\right),
\end{equation}
\noindent
where the derivatives can be described by further PDEs,
\begin{multline}\label{eq:dvxequation}
\frac{\partial \sigma_{x}^2}{\partial t} = -u\cdot \nabla_x \sigma_{x}^2 - v\cdot \nabla_y \sigma_{x}^2 + \nu \nabla^2 \sigma_{x}^2 \\- H\left((t - t_0) - \frac{\norm{P - P_0}_2}{v_{\text{sensor}}}\right) \frac{k_s \sigma_{x}^2}{dt} \sum_{i=0}^{P}{\left[H(r - \norm{\delta\Omega - P_i})\right]} \\+\frac{D_h(u,v)}{dt} + \epsilon_x + E_{k_x}(P),
\end{multline}
\noindent
and
\begin{multline}\label{eq:dvyequation}
\frac{\partial \sigma_{y}^2}{\partial t} = -u\cdot \nabla_x \sigma_{y}^2 - v\cdot \nabla_y \sigma_{y}^2 + \nu \nabla^2 \sigma_{y}^2 \\- H\left((t - t_0) - \frac{\norm{P - P_0}_2}{v_{\text{sensor}}}\right) \frac{k_s \sigma_{y}^2}{dt} \sum_{i=0}^{P}{\left[H(r - \norm{\delta\Omega - P_i})\right]} \\+\frac{D_h(u,v)}{dt} + \epsilon_y + E_{k_y}(P).
\end{multline}
\noindent
In \eqref{eq:dvxequation} and \eqref{eq:dvyequation}, $\nu \in \Re_+$ is a diffusion coefficient, $H(\cdot)$ is the Heaviside step function, used to activate sensing after sufficient time for a sensor travelling at speed $v_{\text{sensor}} \in \Re_+$ to reach a location and to remove the uncertainty tracer in a radius around the sensor position. The sensor effectiveness coefficient $k_s \in [0,1]$ defines how much uncertainty as a proportion of the amount present should be removed by a reading. The variance of the random walk that models turbulent diffusion \citep{Hodgson2019} is described by $\frac{D_h(u,v)}{dt} \rightarrow \Re$ where $D_h(u,v) : \Re \times \Re \rightarrow \Re_+$ is the horizontal diffusion coefficient. An input of uncertainty, $E_{k_x}(P), E_{k_y}(P) \rightarrow \Re$, is a function of the covariance matrix of the time-varying Kalman filter estimations of ${U}_{c}$ and ${U}_{w}$ in horizontal and vertical directions. The terms $\epsilon_x, \epsilon_y \in \Re_+$ are the variance of $U$ from external data sources of ${U}_{c}$ and ${U}_{w}$, or the sample variance of $U$ for that spatio-temporal location.

\section{Optimisation}\label{sec:Optimisation}
The optimisation problem places sensors to minimise uncertainty across the spatio-temporal domain, with prioritised or excluded locations and constraints. Evaluation of $J$ in \eqref{eq:costfunction} is performed upon a co-located grid with forward Euler time stepping. The selection of $E(\boldsymbol{x},t)$, the matrix that weights an area of uncertainty to be minimised, allows sensors to prioritise areas where the information value is high. The weighting matrix is defined by
\begin{subequations}
\begin{align}
\begin{split}
E(\boldsymbol{x},t) &= \frac{1}{k_T} \biggl( k_{P_{\hat{O}_v}} P_{\hat{O}_v}(x_i,y_j) + k_{S_e} S_e(P_{\hat{O}_v}(x_i,y_j),x_i,y_j) \\ &+ k_{P_{\text{DMD}}} P_{\text{DMD}}(x_i,y_j) + k_{\delta\Omega} \biggr),
\end{split}
\end{align}
\begin{equation}
k_T = k_{P_{\hat{O}_v}} + k_{S_e} + k_{P_{\text{DMD}}} + k_{\delta\Omega}
\end{equation}
\end{subequations}
\noindent
where $S_e(P_{\hat{O}_v}(x_i,y_j),x_i,y_j) \rightarrow [0,1]$ is the min-max normalization \citep{Juszczak2000} evaluation of the Shannon entropy \citep{Shannon1948} of $P_{\hat{O}_v}(x_i,y_j)$ in the 3-by-3 neighbourhood of the cell at $(x_i,y_j)$. The total weighting $k_T \in \Re_+$ is the sum of the weighting coefficients $k_{P_{\hat{O}_v}}, k_{S_e}, k_{P_{\text{DMD}}}, k_{\delta\Omega} \in \Re_+$. 

A spatial weighting of critical flow measuring locations $P_{\text{DMD}}(x_i,y_j,t) : \delta \Omega \times \Re_+ \rightarrow [0,1]$, is found through analysis of the Dynamic Mode Decomposition (DMD) model. A truncated to $n_z \in \mathbb{N}$ singular value decomposition of $\mathcal{X}_{k-1|k-1}$ forms the left unitary matrix  $\vec{U} \in \Re^{n_s n_x n_y \times n_z}$, the singular value matrix $\vec{S} \in \Re^{n_z \times n_z}$ and the right unitary matrix $\vec{V^*} \in \Re^{n_z \times k-1}$. The matrix $\vec{U}$ is also a basis function that maps from mode state estimates $\hat{z} \in \Re^{n_z}$ to physical states $\hat{x} = \vec{U} \hat{z}$

Critical locations are calculated in a similar manner to \citep{Annoni2018}. Unlike \cite{Annoni2018}, which utilises a QR factorisation with column pivoting of $\vec{U}^{*}$ to identify $n_z$ points that best sample the modes in $\vec{U}$, an Interpolative Decomposition approximation based on Strong Rank Revealing QR decomposition \citep{Cheng2005} is used instead. See \cite{Martinsson2019} and \cite{KishoreKumar2017} for a detailed comparison and discussion, but in short, QR factorisation resembles
\begin{equation}
\vec{U}^{*}\vec{P} \approx \vec{Q}\vec{R}
\end{equation}
\noindent
where $\vec{P} \in \Re_{+}^{n_z \times n_z}$ is the permutation matrix of columns of $\vec{U}^{*}$ (hence rows of $\vec{U}$) and $\vec{Q} \in \Re^{n_s n_x n_y \times n_s n_x n_y}$ is a unitary matrix, $\vec{R} \in \Re^{n_s n_x n_y \times n_z}$ is an upper-triangular matrix, and $\vec{P}$ is chosen so that the diagonal elements of $\vec{R}$ are non-increasing. The row selections in $\vec{P}$ provides the locations in $\delta\Omega$. It is worth noting that in \cite{Annoni2018} the number of sensors and number of DMD modes must be equal, which is not a limitation acceptable here. The alternative method of the Interpolative Decomposition approximation is described by
\begin{equation}
\vec{U} \approx \vec{K} \vec{U}(\vec{J},:)
\end{equation}
\noindent
where $\vec{K} \in \Re^{n_s n_x n_y \times n_z}$ is a projection matrix and $\vec{J} \in \mathbb{N}^{N_p}$ selects rows for $N_p$ sensors, where each row index maps to a location in $\delta\Omega$. A further difference is that the Interpolative Decomposition approximates $\vec{U}_S$ which is defined by
\begin{equation}
\vec{U}_S = \vec{U} \left(\frac{\vec{S}}{\norm{\vec{S}}_2}\right)^{k_{\text{ID}}}.
\end{equation}
\noindent
Each mode in the unitary matrix $\vec{U}$ is scaled by the corresponding normalised \citep{Juszczak2000} singular value raised to a power coefficient $k_{\text{ID}} \in \Re$ that allows adjustment of the importance of the singular value. For example, $k_{\text{ID}} < 0$ would favour less energetic modes, $k_{\text{ID}} = 0$ would weight all modes equally and $k_{\text{ID}} > 0$ would favour more energetic modes for sampling. A value of $k_{\text{ID}} = 0.5$ is utilised and hence the $N_p$ identified points will be focused on best sampling the more energetic modes of $\vec{U}$.

The spatial weighting of measuring locations $P_{\text{DMD}}$ is formed by assigning scalar values to the $N_p$ identified locations in $\delta\Omega$, where the location closest to an oil presence has a value of $1$ and the further locations have decreasing values according to a normalisation of $\frac{1}{m(N_i)}$, where $N_i : \{\mathbb{N} : i \leq N_p\} \rightarrow \Re^2 \in \delta\Omega$ is the selected critical location and $m(N_i) : \delta\Omega \times \Re_+ \times \Re_+ \rightarrow \Re_+$ is the 2-norm of the Euclidean distance from location $N_i$ to the closest $P_{\hat{O}_v}(x_i,y_j) \ne 0$ in $\delta\Omega$.

\subsection{Moving sensors}
Velocity and position constraints on the sensors also need to be included. Although constraints can be placed on sensors through solver inequalities, this does not produce gradient information that the solver can utilise. There also needs to be sensor guidance in the absence of any oil or uncertainty information, when the sensor is flying over land for example. Constraints are included in the function of interest $J$ by the penalty function $c(\boldsymbol{x},P)$, defined by
\begin{align}\label{eq:functionofinterestvelocity}
\begin{split}
c(\boldsymbol{x},t,\delta\Omega,P,P_0,v_{\text{sensor}}) &= V(P,P_0,v_{\text{sensor}},\delta\Omega) \\ &+ D_m(E(\boldsymbol{x},t)  I(\boldsymbol{x},t),P) \\&+ D_e(\delta\Omega,P).
\end{split}
\end{align}
\noindent
A velocity penalty term $V(P,P_0,v_{\text{sensor}},\delta\Omega) : \delta\Omega \times \Re^{1\times 2 N_p} \times \Re_+ \rightarrow \Re$ maps spatial locations to the euclidean distance to each sensor, though destinations reachable within a given time step incur zero penalty. Descending the derivative $\frac{dV(P,P_0,v_{\text{sensor}},\delta\Omega)}{dP}$, moves sensor positions towards reachable locations.

A forcing term moving sensors towards an area of interest in the absence of other information, $D_m : \Re^{1\times 2 N_p} \times \Re_+ \rightarrow \Re$, is defined for each sensor position and contains the Euclidean distance to the closest region of interest, where $E(\boldsymbol{x},t) q(\boldsymbol{x},t,P) \in \Re_+$. Descending the derivative $\frac{dD_m}{dP}$ moves sensor positions towards the closest region of interest.

The term $D_e : \delta\Omega \times \Re^{1\times 2 N_p} \times \Re_+ \rightarrow \Re$ is introduced to coerce sensors out of an excluded area and is defined for each sensor position as the Euclidean distance to the closest non-excluded area. The derivative $\frac{dD_e}{dP}$ contains the negative of the vector to the nearest non-excluded area for each sensor in the excluded area, with $\frac{dD_e}{dP}$ producing a gradient for sensors to descend towards a permissible area.

\subsection{The adjoint method}
The optimisation problem \eqref{eq:costfunction} is solved through a gradient descent method, using an application of the adjoint method to provide gradient information, in a similar manner to \cite{Funke2014}. The adjoint approach first solves the adjoint equation,
\begin{equation}\label{eq:adjointequation}
\frac{\partial F}{\partial x_{\text{adj}}} ^* \lambda = \frac{\partial J}{\partial x_{\text{adj}}},
\end{equation}
\noindent
where $\frac{\partial F}{\partial x_{\text{adj}}} ^* \in \Re^{n_s n_x n_y k \times n_s n_x n_y k}$ is the jacobian of the finite difference representation of the set of constraint equations \eqref{eq:optracer} to \eqref{eq:optimpos}, $(\cdot)^*$ is the conjugate transpose, $\lambda \in \Re^{n_s n_x n_y k}$ is the adjoint solution, $x_{\text{adj}} \in \Re^{n_s n_x n_y \times k}$ is the state trajectory formed by column stacking the optimisation states and $J$ is the function of interest \eqref{eq:costfunction}. The solving of \eqref{eq:adjointequation} for the adjoint variable $\lambda$ enables calculation of the cost function gradient by 
\begin{equation}
\frac{dJ}{dP} = -\lambda^*\frac{\partial F}{\partial P} + \frac{\partial J}{\partial P}.
\end{equation}
\noindent
The optimal sensor positions are a function of the time-horizon $t_0$ to $t_f$ in \eqref{eq:costfunction}, therefore parallel sequences of adjoint solved optimisations using different time-horizons for each position optimisation produce a sensor path up to a common future time, and the lowest total cost function (formed from the sum of $J$ at each sensor step) is selected.

\subsection{Gradient descent solver}
The initial estimate of sensor placement is at the local maxima of the mean uncertainty across the time step of an uncertainty model run without sensors, ordered from the highest valued maxima to the smallest. The cost function without sensors is described by
\begin{equation}
J_{\text{empty}} = \int _{ t_{ 0 } }^{ t_{ f } }{ \int _{ \delta\Omega  }^{  }{ E(\boldsymbol{x},t)  q(\boldsymbol{x},t)^2 \quad d\delta\Omega  }  } dt.
\end{equation}
\noindent
The local maxima are found through a search of the discrete spatial domain for values higher than their immediate neighbours. The initial sensor positions are defined by
\begin{equation}
P_0 = f_p\left(\frac{J_{\text{empty}}}{t_f - t_0},P_n\right),
\end{equation}
\noindent
where the function $f_p$ finds the highest $P_n$ number of peaks (one for each sensor) and returns their coordinates. Sensor positions descend the gradient each iteration, described by
\begin{equation}\label{eq:gradstep}
\frac{dP}{dn} = \gamma_n \circ \frac{dJ}{dP}_n,
\end{equation}
\noindent
where the step size $\gamma_n : \Re^{1\times 2 N_p} \times \Re_+ \rightarrow \Re$ is found with a backtracking determined line-search using the Armijo-Goldstein condition \citep{Armijo1966,Goldstein1965,Coope1995} for each sensor and $\circ$ denotes the Hadamard product. Gradient descent continues until $\frac{d \text{J}}{dP} < \zeta_g$ where $\zeta_g = 10^{-3}$ is a threshold value, or descent continues to a maximum number of iterations. 

\section{Dynamic Mode Decomposition and State Estimator}\label{sec:DMDAndEstimation}
Sensor data, assumed to be point measurements of states, must be used to estimate the entire environment flow fields and states. Due to the complexity of SCEM and the high numbers of states (e.g 7934400 in the example of Section \ref{sec:SimulationAndResults}) a full-state estimator is infeasible. The work here-in makes use of the Dynamic Mode Decomposition (DMD) \citep{Schmid2014} to form low-order modal models to approximate the dynamics of SCEM, utilising prior work \citep{Annoni2018} to continue a theme of identifying the effective dynamics and changes in those dynamics with new information \citep{Hemati2014}. However, unlike much prior work e.g \cite{Jovanovic2014} and \cite{Brunton2015}, the estimation and accuracy of the DMD model and states is not the primary focus of this research, which instead is to estimate a tracer with non-linear and stochastic dynamics that are partially dependent upon the outputs of the modal model. A Dynamic Mode Decomposition \citep{Schmid2014} of an augmented time-window of the SCEM state trajectory forms a low order (5 modes in the example of Section \ref{sec:SimulationAndResults}) model for state estimation using a time-varying Kalman filter \citep{Chauvin2005} and measured point data. The estimated modal amplitudes are then used to reconstruct an estimate of the states of the higher order SCEM. The reduced order model also predicts a future state trajectory to augment the external data used as inputs for SCEM's prediction. The DMD model is also used to identify critical measuring locations in the flow field, $P_{\text{DMD}}$.

\section{Simulation and results}\label{sec:SimulationAndResults}
This test case is a hypothetical 100 barrel spill of light crude oil near Hong Kong  at 1900 hours on the 8th of January 2019. To provide measurements, 4 mobile sensors capable of measuring oil particles, wind and current velocities arrive 1 hour after the leak begins and stay for 14 hours, de-activating at 0900 hours on the 9th January. The sensors are speed limited to 60 miles-per-hour, model guided sensors are measuring only in 15-minute intervals at point locations while industry sensors have been given the capability to continuously measure while following waypoints. Spill prediction continues to 1900 hours on the 9th of January 2019.

The real simulation, from which sensors measure, utilises data from the Global-Forecast-System (GFS) for wind velocities and Tidetech data for current velocities that include both global circulation currents and tidal flow. The test simulations use the same wind velocity data, but instead use GFS current data that does not include tidal flow which is critical for spill prediction in this region, at the mouth of the Zhujiang river. 

\begin{figure*}
	\centering
	\begin{minipage}{.48\textwidth}
		\centering
		\includegraphics[width=0.95\textwidth]{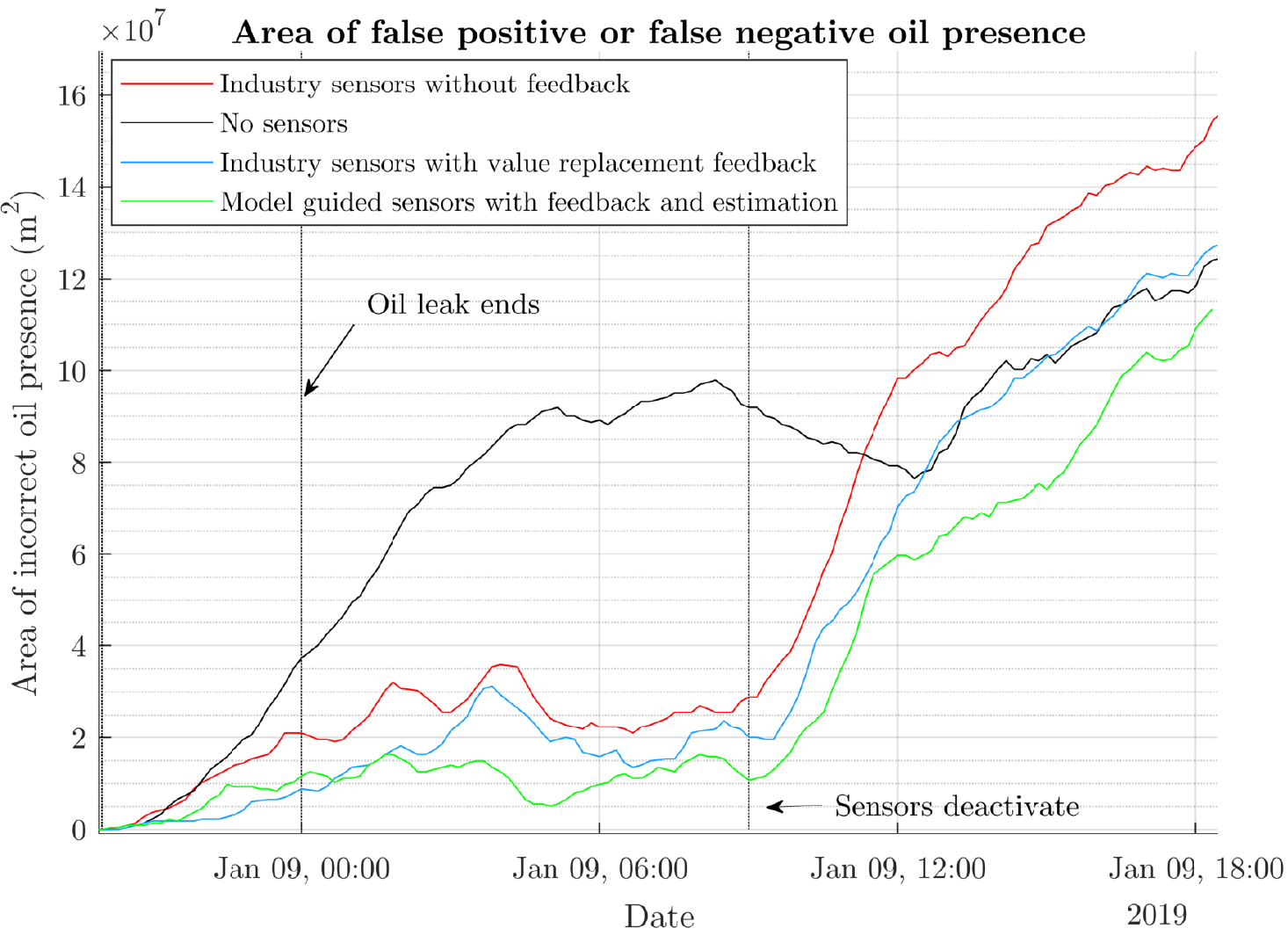}
		\captionsetup{singlelinecheck=off}
		\caption[]{\label{fig:oilerror}
			A comparison of the oil presence error of the simulations. Note how all sensor approaches reduce error by 70\% while sensors are active, with the model guided sensors being approximately 60\% as erroneous as industry standard pathing and continuing to have less error after sensing stops. \newline}
	\end{minipage}
	\hfill
	\begin{minipage}{.48\textwidth}
		\centering
		\includegraphics[width=0.95\textwidth]{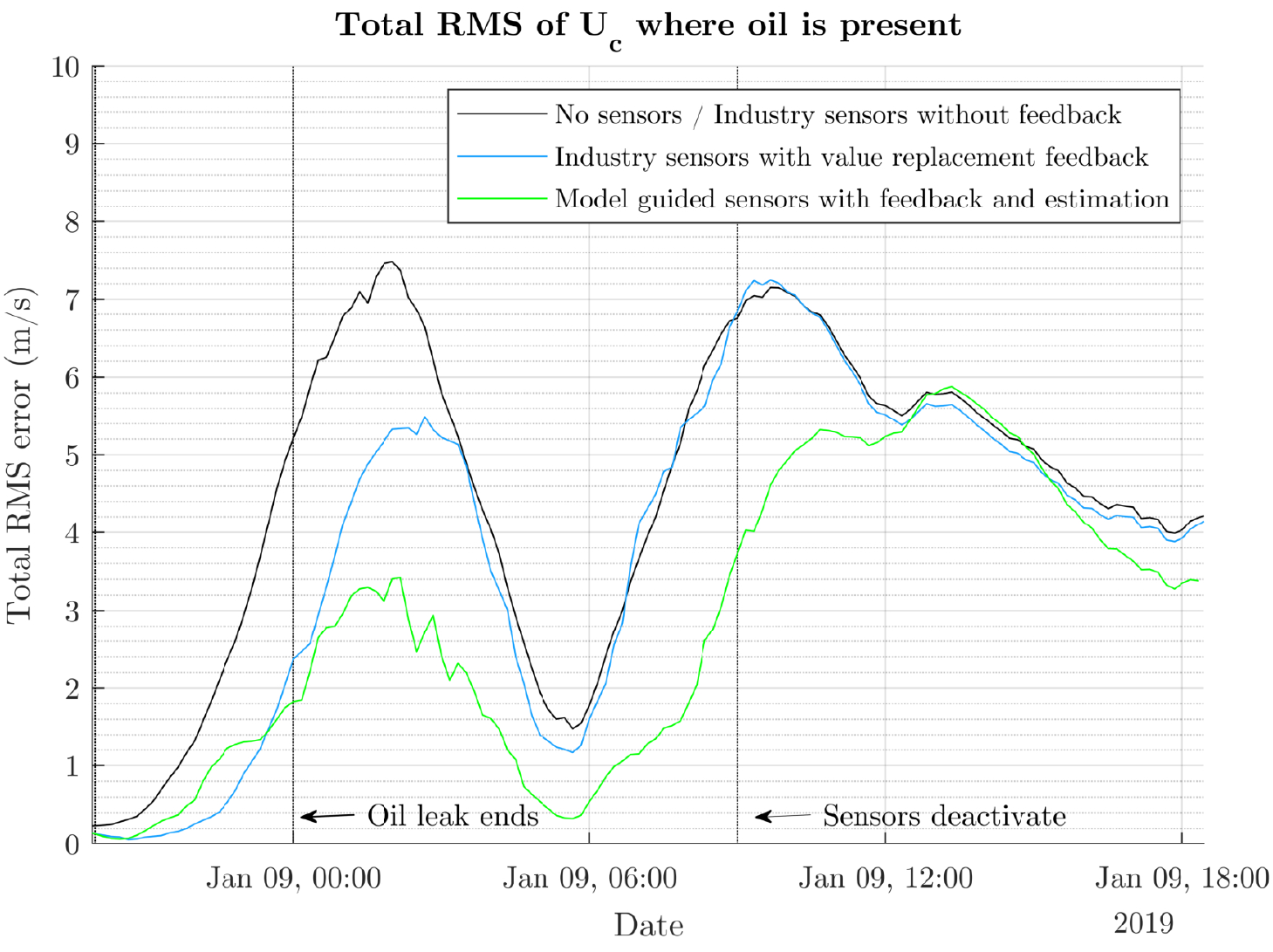}
		\captionsetup{singlelinecheck=off}
		\caption[]{\label{fig:oilerrorcur}
			A comparison of the RMS error of current flow velocity where oil is present. Note the 30\% to 50\% in reduction in error the model guided sensors with feedback and estimation displays over the industry sensors using value replacement, and the continuation of less error after sensing stops. \newline}
	\end{minipage}
\end{figure*}

Industry pathing prescribes a ladder flight path \citep{IPIECA2016,ITOPF2011e} that covers, with a 10\% overlap, where oil is predicted to be by the model. The path plan is split up into sections, one for each sensor, with spacing sufficient to ensure no oil can be missed during flight. Detected oil or clear areas are updated in the model, but velocity and wave spectrum data is only utilised as a value-replacement in the model to reflect the inability of traditional models to utilise measured velocity data in the same manner as SCEM. Figure \ref{fig:oilerror} also includes the error of industry pathing with no velocity feedback, to represent a simple model incapable of modifying external flow data. The ladder flight path is updated every hour to enable sensors to respond to measured oil, with sensors repeating the path at maximum speed for the highest frequency of measurements along the path.

\begin{figure*}
	\centering
	\includegraphics[width=0.55\textwidth]{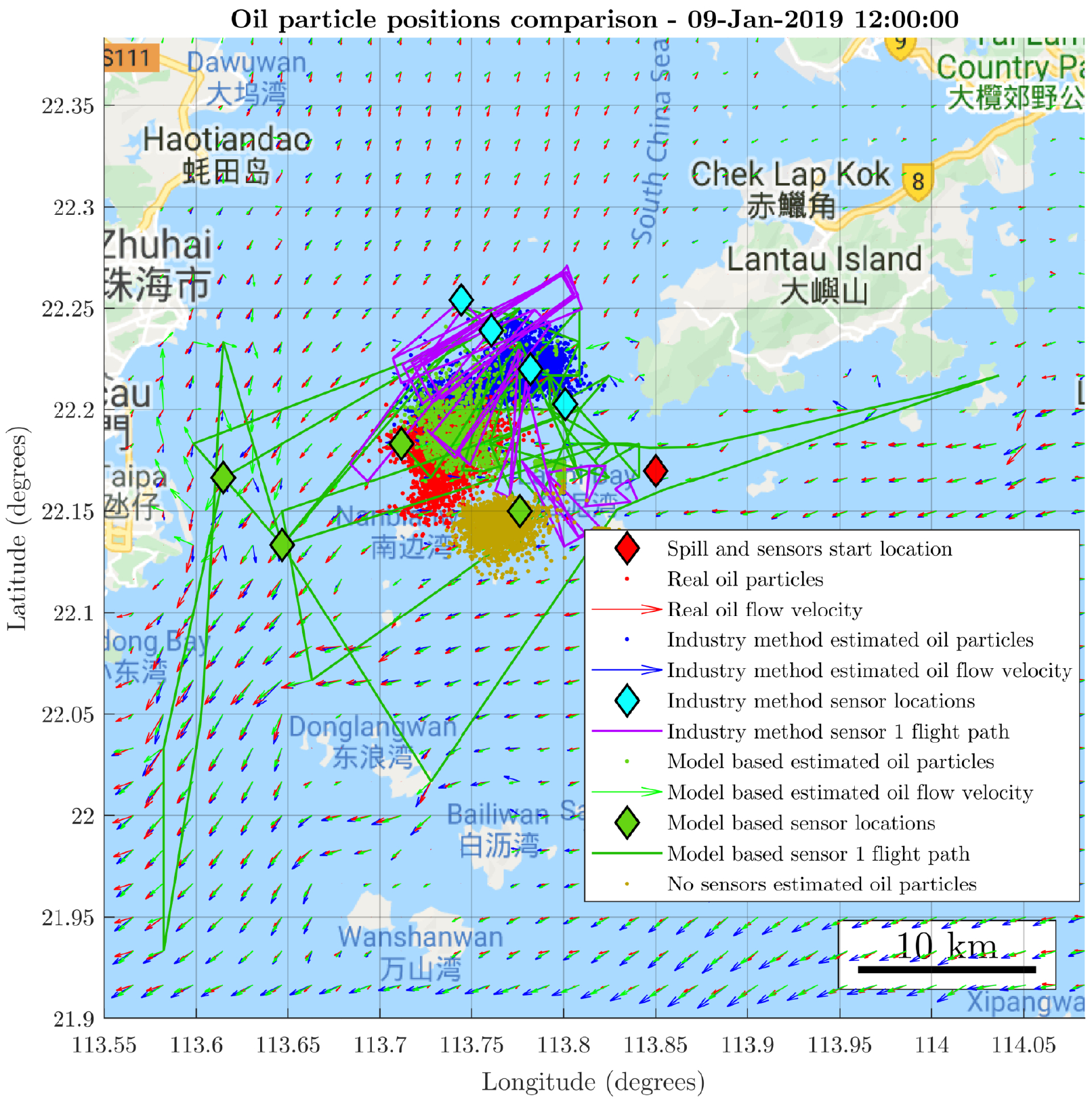}
	\captionsetup{singlelinecheck=off}
	\caption[]{\label{fig:oilcomp}
		A comparison of the simulations at 1200 hours on the 9th January 2019, 15 hours after the initial leak at the indicated spill location, and 3 hours after the 4 sensors have deactivated though their final positions are displayed. Real oil particle locations are displayed in red and are advected by the red velocity field. The simulation using incorrect input data and no sensors has particle locations displayed in gold and the main body of this spill is not within the real spill body. The simulation using the industry method of sensor pathing and feedback, has particle positions and velocity field displayed in dark blue. The main body of the industry spill is 5km to the North East of the real spill. The simulation using model based sensor behaviour and state estimation has the main body of particles (green) within the body of the real spill. Note how the flight path of industry sensor 1 (purple) concentrates over the predicted spill location in an expanding ladder path from the spills initial position, while the flight path of model based sensor 1 (dark green) also flies to crucial velocity measuring locations both up and downstream of the spill, before returning to check the spill.
	}
\end{figure*}

\subsection{Analysis of results}
The simulation using industry pathing and oil particle updates was accurate before sensors were deactivated (see Figure \ref{fig:oilerror}), but once sensors deactivate and the model loses high frequency updates on particle positions, the inaccurate velocity field causes the main body of the industry spill to drift 5km to the North East of the real spill. For the model-based method, after sensors were deactivated the prediction model of SCEM had been sufficiently modified by measured data to produce a more accurate velocity field and maintain accuracy when advecting the particles, with the main body of particles within the real spill location even 3 hours after sensors had deactivated.

Analysing the oil presence error in Figure \ref{fig:oilerror}, model guided sensing with state estimation has a 30\% reduction in the area of incorrect oil presence from the industry method with feedback, or 50\% better than the industry method with only oil information feedback, both when sensors are active and after sensor removal at 0900 hours on the 9th January. After sensor deactivation, the industry method rapidly becomes less accurate as it is still utilising the incorrect input data to predict the spill drift. The model guided sensors have partially corrected the SCEM fluid model to include tidal flow and so while this prediction also loses accuracy after sensors are deactivated, it continues to perform better than using no sensors and the incorrect input data, unlike the industry method. In both Figure \ref{fig:oilerror} and Figure \ref{fig:oilerrorcur} the error of the model guided sensors is slightly above that of industry sensors with value replacement feedback. This is due to the sensors flying further from the spill to measure crucial flow regions and temporarily compromising their update-rate of states local to the spill, though is important for reducing the long-term error.

Although the area covered by the model based flight path of sensor 1 is much greater, as seen in Figure \ref{fig:oilcomp}, the distance moved is actually the same or less than the sensor 1 using the industry method: In the industry method, the sensors fly the ladder path at their maximum speed for the whole time sensors are active, repeating the path and measurements as often as possible before the next ladder path is generated as the update rate is crucial to the accuracy of the industry method. Meanwhile, the model-based method will relocate sensors to optimal positions that may or may not require flying at maximum speed.

\section{Conclusion}\label{sec:Conclusion}
This paper has described a framework for model-based sensor guidance and adaptive monitoring of an oil spill, to guide mobile sensors such as UAVs in gathering information for the support of clean-up operations and incident responses. The framework has demonstrated improvement in monitoring on a test case and a capability for online model adjustment to better predict future spill dynamics. Future work will examine the forms and solution methods of the optimisation problem in \eqref{eq:costfunction} to \eqref{eq:optimvel} and the approaches of system identification and state-estimation used in the framework.

\bibliography{ifacconf_hodgson_arxiv}             

\end{document}